\documentclass{epl}

\title{Fractional power-law susceptibility and specific heat in low temperature insulating state of o-TaS$_{3}$ }
\author{K. Biljakovi\'{c}\inst{1} \and M. Miljak\inst{1} \and D. Stare\v{s}ini\'{c}\inst{1} \and    
J. C. Lasjaunias\inst{2} \and P.Monceau\inst{2} \and H. Berger\inst{3} \and F. Levy\inst{3}}
\institute{
  \inst{1} Institute of Physics, Hr-10 001 Zagreb, P.O.B. 304, Croatia\\
  \inst{2} CRTBT-CNRS, 38042 Grenoble Cedex 9, BP 166, France\\
  \inst{2} IPA, EPFL, 1015 Lausanne, Switzerland
}

\begin{document}

\maketitle

\begin{abstract}
Measurements of the magnetic susceptibility and its anisotropy 
in the quasi-one-dimensional system o-TaS$_{3}$ in its low-T charge 
density wave (CDW) ground state are reported. Both sets of data 
reveal below 40 K an extra paramagnetic contribution obeying 
a power-law temperature dependence $\chi$(T)=AT$^{-0.7}$. The 
fact that the extra term measured previously in specific heat 
in zero field, ascribed to low-energy CDW excitations, also follows 
a power law C$_{LEE}$(0,T)=CT$^{0.3}$, strongly revives the case of 
random exchange spin chains. Introduced impurities (0.5\% Nb) 
only increase the amplitude C, but do not change essentially 
the exponent. Within the two-level system (TLS) model, we estimate 
from the amplitudes A and C that there is one TLS with a spin 
s=1/2 localized on the chain at the lattice site per cca 900 
Ta atoms. We discuss the possibility that it is the charge frozen 
within a soliton-network below the glass transition T$_{g}\sim$40 
K determined recently in this system. 
\end{abstract}

The universality of the low-T thermal properties of amorphous 
materials remains a striking and almost unexplained property. 
Thus below 1 K, glasses, polymers and amorphous materials exhibit 
a roughly linear in T extra-phononic contribution to the specific 
heat C$_{p}$, which have commonly been described through the standard 
model of two-level systems (TLS) intrinsic to structural disorder 
\cite{1}. However it was demonstrated that randomness can also be 
revealed in long range order (LRO) ground state. A new class 
of materials exhibiting low-T universal features of glasses are 
the quasi one-dimensional conductors in the Peierls state characterized 
by a periodic lattice distortion accompanied by a charge density 
wave (CDW) \cite{2}. The model Hamiltonian derived by Fukuyama, Lee 
and Rice \cite{3} comprises an elastic energy term indicating the 
ability of the CDW to be stretched or compressed by pinning centers 
competing with an impurity pinning energy term. The randomness 
of the CDW ground state is reflected in the existence of many 
metastable states and hence there is a strong similarity with 
the X-Y magnet in a random field.

The universality of the disorder property is revealed in an extra-phononic 
contribution to the specific heat with a smooth variation $C_{p}\propto T^{\nu}$ 
in various inorganic quasi-1D CDW compounds \cite{4} with 0.2$<\nu <$0.8. 
The same kind of contribution has been also found in some organic 
quasi 1-d conductors exhibiting a SDW or a spin-Peierls ground 
state, with $\nu$ up to 1 \cite{5}. For all these compounds the 
corresponding LEE excitations demonstrate very peculiar dynamical 
properties which can be understood only if they are weakly coupled 
to the phonon bath, underlying a clear difference from the case 
of conventional glasses \cite{4,5}. In addition, very specific features 
are found for the glass transition evidenced in the SDW state 
\cite{6}, as well as in the CDW state \cite{7}, demonstrating the essential 
role of the screening of the CDW deformations by free carriers 
in the substantial changes in the DW ground state. From this 
point of view, it is very important to better identify the common 
microscopic origin of LEEs and more specifically the basic characteristics 
of the topological disorder, which is naturally expected to be 
related to the DW superstructure.

The fractional-power dependence of $C_{p}\sim T^{\nu}$ is 
usually described through a singularity in the density of states 
$n(E)=n_{o}\cdot E^{\nu-1}$. The same density of state (DOS) can 
yield a singular T-dependence of the magnetic susceptibility, $\chi \sim T^{\nu-1}$. 
The first experimental evidence of this kind of phenomenon has 
been found in complex salts of tetracyanoquinodimethane (TCNQ) 
by Bulaevskii et al. \cite{8}. It has been attributed to the strong 
anisotropy of the electronic structure, which makes the magnetic 
exchange interaction 1-dimensional (with very weak magnetic interactions between 
chains) and leads to the singular DOS. In the 
approach based on the random exchange Heisenberg antiferromagnetic 
chain (REHAC) model a phenomenological DOS has been used, with parameters 
$n_{0}$ and $\nu$ taken from experiment. However, the necessity of a singular distribution of 
random exchange was disproved since even discrete distribution, as in a chain broken into segments or interrupted 
strands, yielded universal REHAC features \cite{19}. Altogether, a lot of theoretical work has 
been devoted to random 1-d magnetic systems (see references in
\cite{9} and \cite{10}).

The CDW system which shows the largest LEE contribution to $C_{p}$ 
among all DW systems \cite{4,11} is o-TaS$_{3}$. It exhibits a Peierls 
transition at T$_{P}$=218 K, below which a CDW develops and turns 
into a low-T insulating ground state \cite{2}. The exponent $\nu \sim 0.3$ 
of $C_{LEE} \propto T^{\nu}$ was low enough for investigation 
of the possible contribution of the corresponding DOS to the 
susceptibility as $\chi_{LEE} \propto T^{\nu-1}$. In 
order to establish the correspondence between the two thermodynamic 
quantities, and add some more information of the low temperature 
state, we have performed measurements of the susceptibility and 
of the susceptibility anisotropy in a wide T-range, from 4 K 
to 300 K.

Magnetic susceptibility $\chi$ was measured by the Faraday method and 
the susceptibility anisotropy $\Delta \chi$ by a home made highly sensitive quartz 
torque magnetometer \cite{12}. The sensitivity of $\Delta \chi$ measurement 
is about 10$^{-12}$emu/g -- for a sample mass of $\sim$10 mg. 
The o-TaS$_{3}$ sample was mounted in a strain free manner, without 
grease or glue, between two thin quartz plates held by a light 
quartz spring. Note that the sample used for magnetic measurements 
(m=7 mg) was from the same batch as the sample used in the specific 
heat experiment. It looks like a carpet consisting of a bunch 
of needles all grown in the same chain direction. The pair of 
crystal axes perpendicular to chains are most probably oriented 
at random since no anisotropy was observed in this plane in the 
whole T-range.

\begin{figure}[t]
\onefigure[height=6cm]{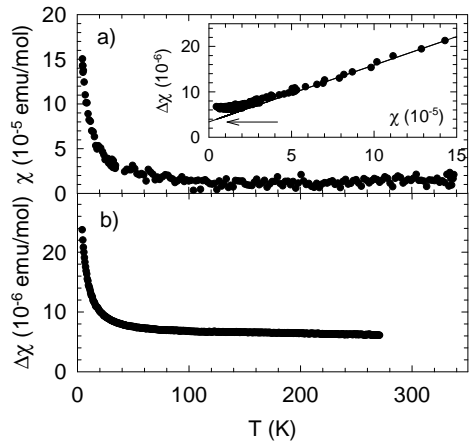}
\caption{$\chi$ (a) and $\Delta \chi$  
(b) versus temperature for o-TaS$_{3}$ both show low-T upturn. The 
inset in (a) demonstrates the linear correlation between both 
sets of data and the possible estimation of the correction value 
indicated by the arrow.}
\label{f.1}
\end{figure}

As seen in fig.~\ref{f.1}, both measurements, $\chi$ (fig.~\ref{f.1}a) 
and surprisingly $\Delta \chi$  (fig.~\ref{f.1}b) reveal a relatively large low-T upturn. 
Judging from the similar 
T-dependence of $\chi$  and $\Delta \chi$ and from the fact 
that the absolute value of $\Delta \chi$ is up to about 10\% of 
$\chi$, $\Delta \chi$ seems to show up via electron 
spin g-factor anisotropy. Inset in fig.~\ref{f.1}a displays the correlation 
between $\chi$ and $\Delta \chi$, showing their proportionality 
in the whole T-range, which brings the verification that the 
assumed source of the anisotropy originates from the g-factor 
anisotropy. Furthermore, the absence of any break in the slope 
proves that the low-T upturn is caused by the same spin carrying 
magnetic species; those also giving the intrinsic paramagnetism 
dominant above 40 K. This high temperature behaviour, unexpected 
for CDW formation, will be discussed elsewhere. 

\begin{figure}
\onefigure[height=6cm]{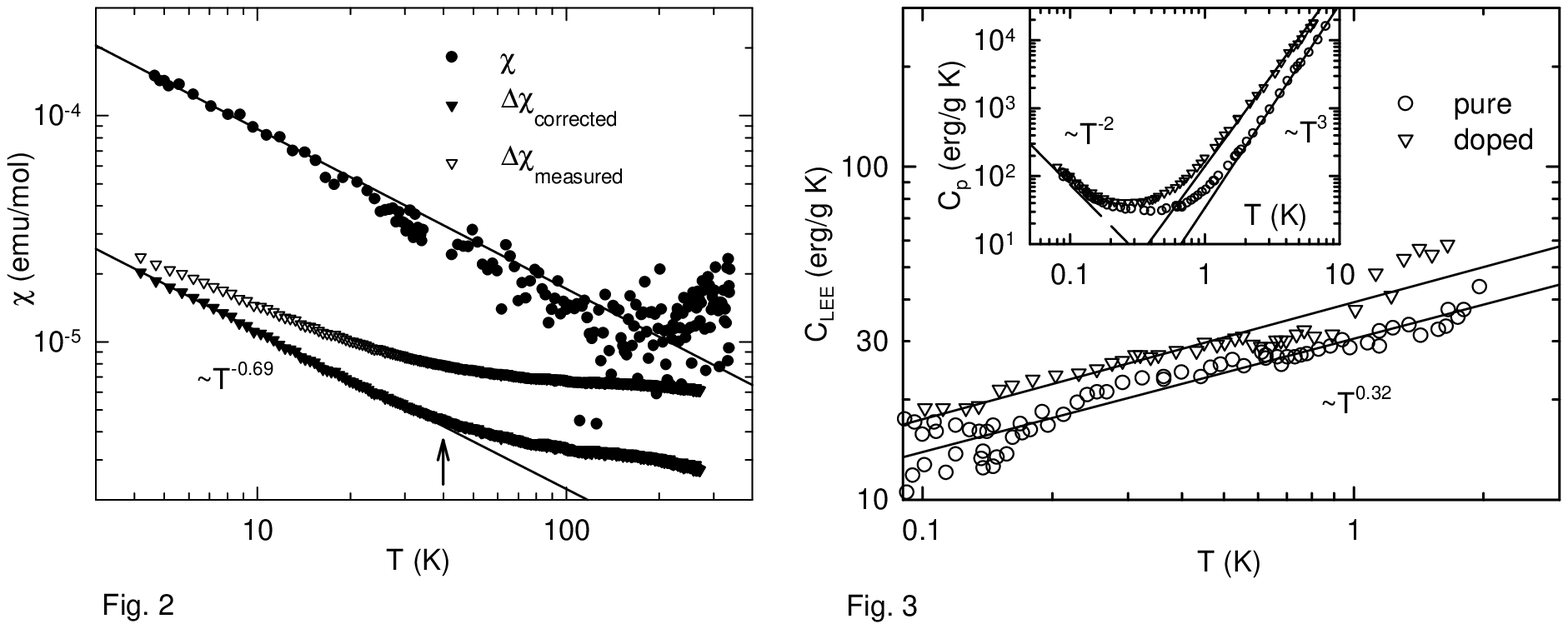}
\caption{Log-log plots of $\chi$ and $\Delta \chi$
versus T exhibit the same power-law dependence after the correction 
(explained in the text) is done. The arrow points to J$\sim $T$_{g}\sim$ 
40 K \cite{8}.}
\label{f.2}

\caption{Heat capacity of pure and Nb-doped (0.5\% nominal) o-TaS$_{3}$ 
between 0. 1 K and 7 K,  analysed following eq.~(\ref{e.1}). 
In the inset is shown the total $C_{p}$. Main figure shows power law contributions 
of the LEEs obtained after subtraction of T$^{-2}$ and C$_{l}$ contributions.}
\label{f.3}
\end{figure}

It should be noticed that the small intercept on $\Delta \chi$
and/or $\chi$ axes for $T\rightarrow 0$ shows that either 
the corrections made to the measured $\chi$ (tabulated 
Pascal constants) may not be correct, or some T-independent contribution 
to $\Delta \chi$ is present. The importance of the proper correction 
is clearly demonstrated in fig.~\ref{f.2}, which displays $\chi$
and $\Delta \chi$ versus temperature in a log-log plot. It shows 
that both measurements can fit the power law $A\cdot T^{\nu-1}$, 
but some corrections should be made to achieve the same exponent \footnote{In 
the course of the intrinsic susceptibility evaluation where relatively 
large low temperature upturn has to be subtracted from the small 
susceptibility values, the choice of the subtracting procedure 
becomes delicate. We find that between the three possibilities 
of the low temperature upturn description: the Curie law $C\cdot T^{-1}$, 
the Curie-Weiss law and the $A\cdot T^{\nu-1}$ power law, the power 
law appears to be the most appropriate one.}. The appropriate correcting values should be within the limits 
of a given intercept on $\chi$ and/or $\Delta \chi$ axes 
displayed in the inset of fig.~\ref{f.1}a. One way is to subtract only 
the small intercept on the $\Delta \chi$ axis from the 
measured anisotropy, i.e. -3.4$\cdot $ 10$^{-6}$ emu/mole marked 
by the arrow, what results both, in the linear correlation and 
the exponent $\nu-1$=-0.694 equal to that one of $\chi$
(-0.696). 

There are a few important conclusions which can be drawn from 
these results. {\underline {First}}, the fact that $\Delta \chi$ in this 
weak paramagnetic material also exhibits low temperature upturn, 
indicates the presence of the paramagnetic ``impurities'' \textit{located 
on the lattice sites}. Otherwise, as in most cases when paramagnetic 
impurities are at random, the contributions to the measured $\Delta \chi$
caused by the g-factor anisotropy in average would cancel out. 
{\underline {Second}}, $\Delta \chi$ measurements, beeing more sensitive,
 enable us to define more precisely the exponent $\alpha=\nu-1$ and also the T-range of the power law fit which 
reflects the range of the exchange interaction $J_{L}$ between localized 
magnetic species. In the absence of downwards deviation of the 
anisotropy from the power-law up to 30 K, we can estimate the 
lower limit of the corresponding magnetic interaction at low-T: 
$J_{L}\sim$ 30-40 K \cite{12}. Unfortunately, without the exact ESR g-factor values 
$\Delta \chi$ measurements cannot be used for the estimation 
of the number of paramagnetic "impurities'' contained in the 
power law prefactor A. {\underline {Third}} point is the estimation 
of the concentration of the paramagnetic "impurities''. There 
are few approaches and regardless the applied model \cite{9,12}, the 
variation is within 50\%. At this place, before entering a discussion 
about the origin of these magnetic species, we choose the extreme 
case -- "impurities'' being free spins. By assuming a Curie law 
with a constant $C_{c}=A=4.3 10^{-4}$K$^{-0.3}$ emu/mol from $\chi$(T$<$4 K) data
 we obtain the spin concentration c=1.15$\cdot $ 10$^{-3}$ mol $^{-1}$. 
Within the purely 1-d REHAC model, it yields linear concentration 
of 3.3 10$^{4}$ cm$^{-1}$ per chain. It means that the spins of s=1/2 
break the Ta-chain into segments of cca 900 Ta atoms.

The heat capacity was measured in a dilution refrigerator by 
means of a conventional heat-pulse transient technique \cite{5,11}. 
When the regular phonon contribution, generally in T$^{3}$, and a "hyperfine" 
contribution in T$^{-2}$ are subtracted, an excess specific heat 
remains below 1K the T-dependence of which shows a power law 
$C\cdot T^{\nu}$ \cite{11}. The total heat capacity for pure and Nb-doped 
o-TaS$_{3}$ samples is reported in inset of fig.~\ref{f.3}. Starting from 
a common lowest-T hyperfine contribution $C_{h}\cdot T^{-2}$, the heat capacity 
of the doped sample shows a larger value in the upper T-range, 
in particular above T$\sim$2 K where $C_{p}$ is dominated by 
the lattice contribution $C_{l}$. We believe that this enhancement 
is due to the strong defective effect (point-defects) induced 
by the atomic substitution of the Ta atoms by lighter isoelectronic 
Nb atoms, which in particular results in variation of $C_{l}$(T) slower 
than cubic.

We analysed $C_{p}$ in the investigated T-range as the sum of three 
contributions. The two first are originating from $C_{LEE}$ - decomposed 
into the hyperfine term $C_{h}\cdot T^{-2}$ and the power-law $T^{\nu}$ \cite{14} 
and the third one is of vibrational origin:

\begin{equation}
\label{e.1}
C_{p} = C_{h}\cdot T^{-2}+C\cdot T^{\nu}+C_{l}
\end{equation}

\begin{table}
\caption{Parameters of the fits of C$_{p}$ experimental data to the eq.~(\ref{e.1}) for pure and 0.5\% Nb-doped TaS$_{3}$.}
\label{t.1}
\begin{center}
\begin{tabular}{ccccc}
o-TaS$_{3}$ & C$_{h}$ (erg K/g) & C$_{l}$ (erg/g K) & C (erg/g K$^{1.32}$) & $\nu$\\
pure & 0.75 & 34 T$^{3}$ & 30 & 0.32 $\pm$ 0.02\\
Nb doped & 0.75 & 140 T$^{2.65}$ & 40 & 0.32 $\pm$ 0.02
\end{tabular}
\end{center}
\end{table}

The parameters of fits for both curves are given in table~\ref{t.1} .

The main result of this analysis is the increase of the amplitude C of the 
power law contribution in the doped sample by cca 30\%, 
whereas the exponent is preserved, as demonstrated in fig.~\ref{f.3}. It is known from our previous work 
that the exponent $\nu$ in the power law contribution is a 
very specific characteristics of the investigated CDW system 
\cite{4}, but it does not change for different sample batches or even 
for doping \cite{15}, hinting to an intrinsic property of each ground 
state. The same has been also found in REHAC systems with similar 
consequences on doping \cite{9} (note that we do not discuss the irradiation 
effect).

As we noticed, this power law dependence of the specific heat 
and magnetic susceptibility can be treated through the phenomenological 
DOS, n(E)=n$_{0}$E$^{\nu-1}$, with two adjustable parameters. 
The calculation of the DOS depends slightly on the type of excitations 
and the difference appears only in the amplitude n$_{0}$. In order 
to calculate the DOS of the LEE, we start with the hypothesis 
generally admitted \cite{16} that CDW metastable states correspond 
to TLS-type excitations rather than harmonic oscillators. Indeed 
the possibility of anharmonicity, in particular if the LEE are 
related to the pinning centers, is better described by the general 
anharmonic character of the double well potential of the TLS 
\cite{17,18} in comparison to the single well harmonic potential. 
Within this frame, we have calculated the number of TLS

\begin{equation}
\label{e.2}
n=\int n(E)dE=1.05\cdot 10^{34} \int\nolimits_{0.1\;K}^{40\;K}E^{-0.68} dE  
\end{equation}

n(E) expressed in units states/erg cm$^{3}$, over the energy range 
0.1 K to 40 K (note that E=2.5 k$_{B}$T is the dominant energy splitting 
contributing to C$_{p}$ measured at T). The upper limit corresponds 
(for being consistent with the susceptibility) to the low-T magnetic 
interaction J$_{L}$(=k$_{B}$T) indicated in fig.~\ref{f.2}. Therefrom we can 
get an estimation of the concentration of the LEE by comparison 
to N=6.021$\cdot $ 10$^{23}$ structural units defined by the chemical 
formula (with M=277 g/mol), being 1030 ppm\footnote{For comparison, 
note that the integration up to E=30 K yields a concentration of 920 ppm.}, very close to the value 
1150 ppm obtained from the susceptibility using Curie law! So, 
each TLS excitation is related to one spin s=1/2.

\textit{What is the real nature of these excitations?} Overall features 
resemble the widely investigated REHAC phenomenon with spin s=1/2 \cite{8,19,9} 
demonstrating ``standard'' fractional-power law thermodynamical 
properties found in a wide family of organic charge transfer 
salts. Unfortunately, the nature of the elementary excitations of REHAC 
has not yet been elucidated, nor is it known what is their degree 
of localization and how it is related to the distribution of 
magnetic interaction \cite{9}. In the following we intend to propose 
the real microscopic picture of these excitations in o-TaS$_{3}$, 
to the best of our knowledge the first inorganic system demonstrating 
REHAC properties.

We notice first that o-TaS$_{3}$ exhibits very close exponents $\nu$ 
and $\alpha$, and even the same effects on doping 
as REHAC, but the great difference appears in the amplitudes 
of the investigated thermodynamical properties and the corresponding 
number of paramagnetic "impurities'', as we provisionally named 
these low-energy elementary excitations. The REHAC contribution 
to the specific heat of mostly investigated Q(TCNQ)$_{2}$ \cite{9} is 
C$_{spin}$=23.6 T$^{0.18}$ mJ/molK, 30 times larger at 1 K than in o-TaS$_{3}$ 
(with comparable lattice contribution)! Similarly, the estimated number of spins responsible 
for the low-T magnetic properties was found to be in the range 
from 3\% to 10\% \cite{9,19}, more than 30 times larger than in o-TaS$_{3}$! 
We might say that o-TaS$_{3}$ is an inorganic REHAC system in a 
highly ``diluted'' limit of paramagnetic ``impurities''. Those 
impurities are spins of s=1/2 sparsely placed on the lattice 
sites cca every 900 Ta distances or 3000 {\AA}, interacting with 
J$_{L}\sim $ 30-40 K, which give rise to the typical REHAC 
behaviour.

The basic characteristics of REHAC - the disorder 
and localization - are inherent to the CDW ground state; so it 
is not surprising that both length scale and energy scale obtained 
from the REHAC behaviour are deeply related to the CDW ground 
state of o-TaS$_{3}$. In fact for being more explicit, we believe 
that it is one additional manifestation of the change within 
the CDW ground state occurring at low-T. We have recently demonstrated that the low temperature ground 
state of o-TaS$_{3}$ is reached through the glass transition on 
the level of the CDW superstructure \cite{7}. We have related it to 
the freezing of dynamics of CDW phase domains due to the Coulomb 
hardening in the absence of free carrier screening. The estimate 
of the critical free carrier density n$_{e}$(T$_{g}$) which leads to 
freezing, to be about 10$^{15}$ e/cm$^{3}$, is consistent with charge 
frozen at T$_{g}$ as obtained from thermally stimulated discharge 
\cite{21}. The corresponding volume of 10$^{-15}$ cm$^{3}$ per free carrier 
is close to the estimates of the phase coherence volume, and 
it might be at the origin of the twinkling domains seen in TEM \cite{22} 
which have just the right size ($\sim$ 4000$\times$300$\times$300 
{\AA}$^{3}\approx$4$\cdot $ 10$^{-16}$ cm$^{3}$). 

Remaining degrees of freedom below T$_{g}$ have been associated 
with topological defects of phase such as solitons, domain walls or 
dislocation loops. REHAC behaviour, which is essentially 
related to 1-dimensionality, can help in further understanding 
of LEE at low temperatures. Recent analysis of the properties 
of CDW in the low-T insulating phase \cite{23} has shown that soliton-like 
defects on single chain, which perturb very weakly the phase 
on adjacent chains, can exist due to nonlinear screening. In 
typical conditions easily met in insulating CDW systems such 
solitons would change local value of electrochemical potential 
allowing creation of strongly 1-d metallic islands. Slight overlap 
of these islands can explain, for instance, enhanced conductivity 
observed (only) in the chain direction \cite{24}. More important for 
us, these metallic islands can give localized unpaired spins leading to the REHAC behaviour due to the inherent strong 
anisotropy. We should add that this is not the single possible 
explanation, as there exist more exotic species of solitions 
that can have spin \textit{per se} \cite{25}. 

There should exist some fine tuning mechanism coming from the 
interaction with underlying lattice responsible for so specific 
positions of the REHAC excitations. A strong influence of the 
CDW on the lattice has been demonstrated in o-TaS$_{3}$ \cite{26}. The 
glass transition itself leaves some fingerprints on the phonon 
dynamics \cite{13}. STM studies showed, that in addition to very complicated 
structure of o-TaS$_{3}$, there are two CDW maxima per unit cell 
along the b$_{0}$ direction which are strongly correlated, so that 
they respond with a single Peierls transition. As there is no 
chain staggering, the CDW maxima are adjacent to each other and 
the Coulomb repulsion encourages the uniform distribution of 
the CDW within the unit cell (i.e. they are not localized on 
any particular chains) \cite{27}. It is very probable that the origin 
of the preferred position of the REHAC elementary excitation 
are the CDW dynamical defects, as sudden sliding of CDW for one 
lattice spacing in the chain direction has been observed even 
at room T \cite{27}. In this very delicate entanglement of the lattice 
and CDW(s), it might be that the lattice tries to stagger the 
chains in order to reduce the Coulomb repulsion. On approaching 
the glass transition the Coulomb hardening of the CDW forces 
the CDW wave vector to fit the commensurate value q$_{2k_{F}}$=1/4 
c$_{0}$. Hence, the solitons, being the defects of the CDW superstructure, 
tend to have preferential positions. Finally, also the structural 
defects might be dragged by the CDW and frozen together on the 
edges of the Lee-Rice domains, consistent with our previous discussion.

In conclusion, we would like to stress the importance of the 
detailed specification of low-T CDW excitations yielding the 
characteristic REHAC behaviour, equally from the experimental 
point of view, as well as theoretical one. Charge-spin mixing 
properties are essentially important as it is known that in CDW 
the soliton generates a localized region of a SDW and a net spin 
(and vice versa) \cite{25}. The problem of various kinds of solitons, 
self trapping and/or self doping seems to be a central point. 
Some further comparison with "classical'' REHAC behaviour, especially 
concerning the dynamical manifestations, might give some answers 
on the relevance of the lengths of correlations (LRO/SRO) in 
the corresponding ground states.

\end{document}